\documentclass[12pt,preprint]{aastex}

 \newcommand{\SST}{{\it SST}}
 \newcommand{\kms}{km~s$^{-1}$}
 
 \newcommand{\HII}{\ion{H}{2}}
 \newcommand{\cmmb}{cm$^{-2}$}
 
 \newcommand{\tco}{$^{12}$CO}

\shorttitle{Bowshock B Star}

\begin{document}

\slugcomment{v. 2011 December 26}

\title{IRAS03063+5735: A Bowshock Nebula Powered by an Early B Star}

\author{Henry A. Kobulnicky, Michael J. Lundquist, Anirban  Bhattacharjee}
\affil{Department of Physics \& Astronomy \\  1000 E. University \\
University of Wyoming \\ Laramie, WY 82071 , USA
\\ Electronic Mail: chipk@uwyo.edu,mlundqui@uwyo.edu,abhattac@uwyo.edu}

\author{C. R. Kerton}
\affil{Department of Physics and Astronomy \\ Iowa State University \\ Ames, IA 50011, USA
\\ Electronic Mail: kerton@iastate.edu}

%\vskip 1.cm

\begin{abstract}

Mid-infrared images from the {\it Spitzer Space Telescope} Galactic
Legacy Infrared MidPlane Survey Extraordinaire program reveal that the
infrared source IRAS~03063+5735 is a bowshock nebula produced by an
early B star, 2MASS~03101044+5747035.  We present new optical spectra
of this star, classify it as a B1.5~V, and determine a probable
association with a molecular cloud complex at $V_{LSR}$=$-38$ -- $-42$
\kms\ in the outer Galaxy near $\ell=140.59^\circ$, $b=-0.250$\degr.
On the basis of spectroscopic parallax, we estimate a distance of
4.0$\pm$1~kpc to both the bowshock nebula and the molecular complex.
One plausible scenario is that this a high-velocity runaway star
impinging upon a molecular cloud.  We identify the \HII\ region and
stellar cluster associated with IRAS~03064+5638 at a projected
distance of 64 pc as one plausible birth site.  The spectrophotometric
distance and linkage to a molecular feature provides another piece of
data helping to secure the ill-determined rotation curve in the outer
Galaxy.  As a by-product of spectral typing this star, we present
empirical spectral diagnostic diagrams suitable for approximate
spectral classification of O and B stars using He lines in the
little-used yellow-red portion of the optical spectrum.

\end{abstract}


\section{Introduction} 

Short-lived massive stars typically lie within the OB associations in
which they were born. However, a significant fraction (17\%--50\%)
having high (30--200 \kms) space velocities wander far from such
associations and are known as ``runaways'' \citep{blaauw, stone82,
conti77, gies86}.  While many runaways are detected  directly on the
basis of large radial velocities or high proper motions, others are
inferred by the presence of a tell-tale bowshock produced by their
supersonic motion through the surrounding medium.   \citet{vanb88},
\citet{vanb95}, and \citet{noriega97} used \emph{Infrared
Astronomical Satellite} (\emph{IRAS}) images to locate arc-shaped
bowshock features associated with high-velocity O stars.   Bowshocks
associated with high-velocity O stars have been detected more easily 
using the recent generation of mid-infrared surveys with higher
angular resolution performed by the \emph{Midcourse Space Experiment}
(\emph{MSX}) and the \emph{Spitzer Space Telescope} (\emph{SST}).  
Recent studies of bowshocks around massive stars  include
\citet{brown05},\citet{comeron07}, \citet{gvar08}, and
\citet{povich}.  In \citet{kobulnicky10} we used the \emph{SST}
Cygnus-X Legacy Survey data to identify several probable runaway
stars and their bowshocks in the vicinity of the Cygnus OB2
association while introducing  a novel technique to exploit bowshock
physics in order to estimate mass-loss rates from massive stars.  

In this contribution we report the discovery that infrared source 
IRAS~03063+5735 is a bowshock nebula associated with an early B star
in the outer Galaxy near $\ell=140.59$\degr, $b=-0.250$ \degr.  Data
from the \emph{SST} Galactic Legacy Infrared MidPlane Survey
Extraordinaire 360 (GLIMPSE 360) mid-infrared survey reveals the
bowshock morphology for the first time.   We use optical spectroscopy
from the Wyoming Infrared Observatory (WIRO)  2.3~m telescope to
determine a spectral type and radial velocity.  These data, in turn,
enable an estimate for the distance to the star and an associated 
molecular cloud complex.  

\section{Data}

\subsection{Discovery and Infrared Imaging Data}

\begin{figure}
\plotone{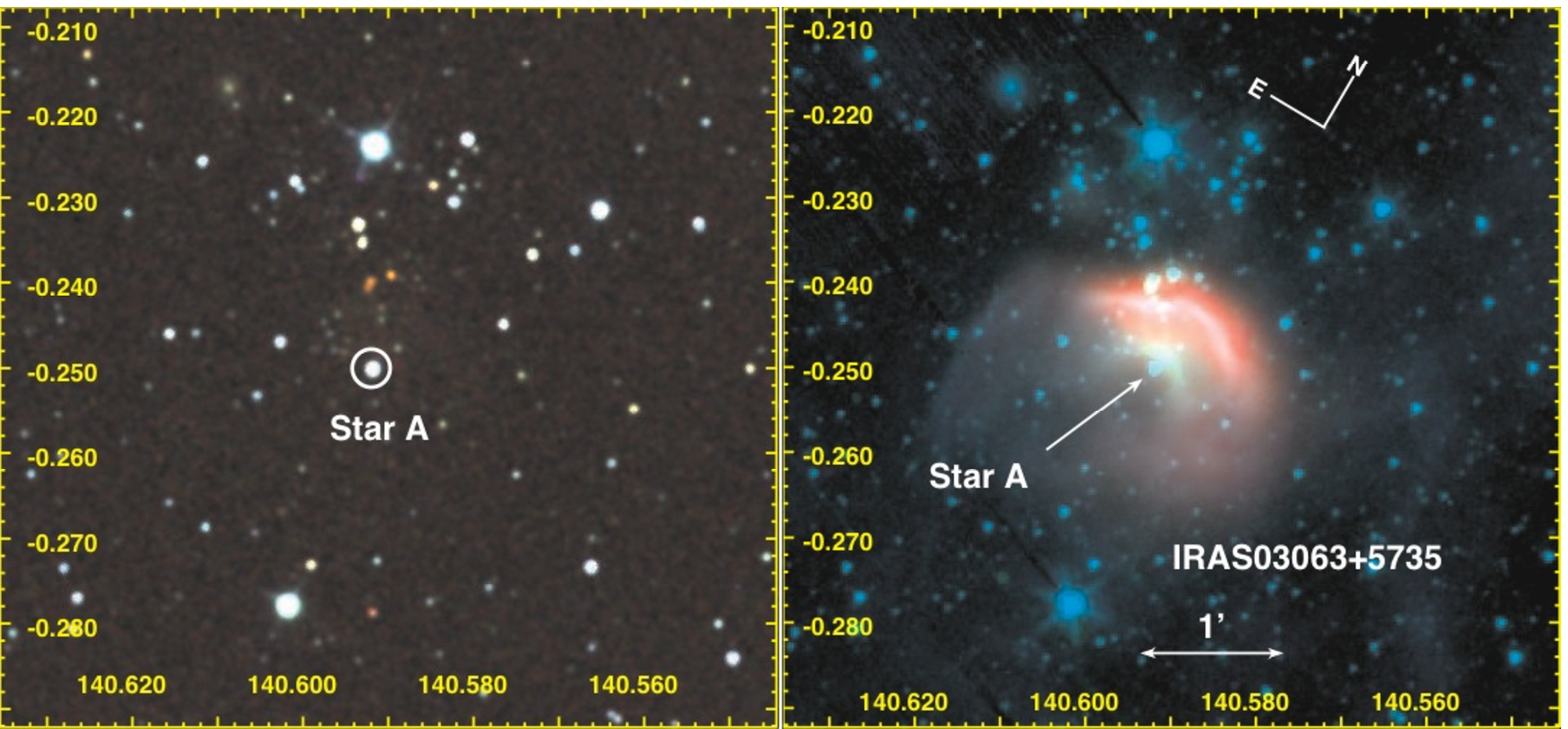}

\caption{(right panel) {\SST} three-color image in Galactic
coordinates of the  IRAS~03063+5735 region with \SST\  [3.6] in blue,
[4.5] in green, and {\it WISE}  W3 12 $\mu$m in red, revealing the double
bowshock morphology of the nebula.  (left panel)  2MASS J/H/K image
of the same field in blue/green/red, respectively. The star
2MASS03101044+5747035 (Star A) lies at the center of the nebula and
can be seen in both images.   \label{discovery} }

\end{figure}

GLIMPSE~360 is a \SST\ warm mission legacy program survey of the
outer Galaxy in the [3.6] and [4.5] $\mu$m bandpasses (bands 1 \& 2)
of the Infrared Array Camera \citep[IRAC][]{fa04}.   As part of a
systematic study of ``intermediate-mass''star-forming  regions where
stars up to $\sim$B0 are being formed, our team has been
investigating IRAS sources having colors  characteristic of soft UV
radiation fields from early-to-mid B stars.     IRAS~03063+5735 was
initially selected as having \emph{IRAS} colors indicating a such a
soft radiation field similar to Galactic star forming regions hosting
intermediate-mass stars \citep{kerton02,arvidsson10}.\footnote{As of
this writing, the  SIMBAD database lists it as a galaxy, presumably
on the basis of its infrared colors, which include some star-forming
galaxies.}   Our visual examination of IRAS~03063+5735 in GLIMPSE~360
images  revealed a striking bilateral symmetric bowshock morphology
at [3.6] and [4.5].  Figure~\ref{discovery} (right panel) shows a
color representation of the nebula with the \SST\ [3.6], [4.5], and
\emph{Wide-Field Infrared Survey Explorer}
\cite[\emph{WISE};][]{wright} band W3 [12.1] $\mu$m image in blue,
green, and red, respectively.  The nebula has a diameter of
$\sim$2\arcmin. A star, 2MASS~03101044+5747035 (hereafter Star A),
lies near its center and along the axis of symmetry, making it the
probable progenitor star.  The left panel shows the 2MASS
\citep{2MASS} $J/H/K_s$ image in blue/green/red, respectively.  Note
the presence of two infrared-bright rims in the mid-IR images.  The
first, with an apsis located $\sim$7\arcsec\ from Star A, is
yellow-green in color, indicating relatively strong emission in the
[4.5] (green) band.  Diffuse emission in this band is typically
dominated by shocked molecular hydrogen often seen in outflows from
young stellar objects \citep{reach06,davis,cyg}, although the
Brackett $\alpha$ line is also in this bandpass.  The second, with
apsis $\sim$27\arcsec\ from Star A, is relatively brighter in the
[3.6] and [12.1] bands, giving it a reddish-white hue in
Figure~\ref{discovery}, but it is present at [4.5] as well.  The
[3.6] band contains broad polycyclic aromatic hydrocarbon (PAH)
features while the [12.1] band contains both PAH features and a
contribution from hot dust grains.

\subsection{WIRO Optical Spectroscopy}

We obtained optical longslit spectra of Star A at the WIRO 2.3~m
telescope on the nights of 2011 October 23 \& 27 using the 900 l/mm
grating in first order over the wavelength range 4300 -- 7300 \AA.  A
total of seven 600-second exposures were obtained with a 2\arcsec\
$\times$ 150\arcsec\ slit in  $\sim$2\arcsec\ seeing with an
east-west slit orientation on 23 October and a north-south slit on 27
October.   Spectra were wavelength calibrated using CuAr lamp
exposures obtained after each exposure of Star A.   The wavelength
calibration has an rms of 0.05 \AA\ and the spectra have a resolution
of 3.9 \AA\ FWHM based on CuAr lines.  Individual spectral exposures
were reduced using standard flat fielding and wavelength
rectification procedures before being transformed to the Local
Standard of Rest (LSR) velocity frame and combined. 

We also obtained single 5--10 s exposures of 22 bright B-type dwarfs
and giants from the Smithsonian Astrophysical Observatory Star
Catalog  for use in calibrating the spectral type of Star A.  The
spectra  were reduced and calibrated in the same manner as described
above.   Table~\ref{lines} lists these stars along with equivalent
width measurements for prominent H and He lines.  These spectra have
typical signal-to-noise ratios (S/N) exceeding 200:1 in the middle of
the spectral range observed.

% this is a comment

\section{Analysis}

\subsection{Optical Spectra and Spectral Type of Star A}

\begin{figure}
\plotone{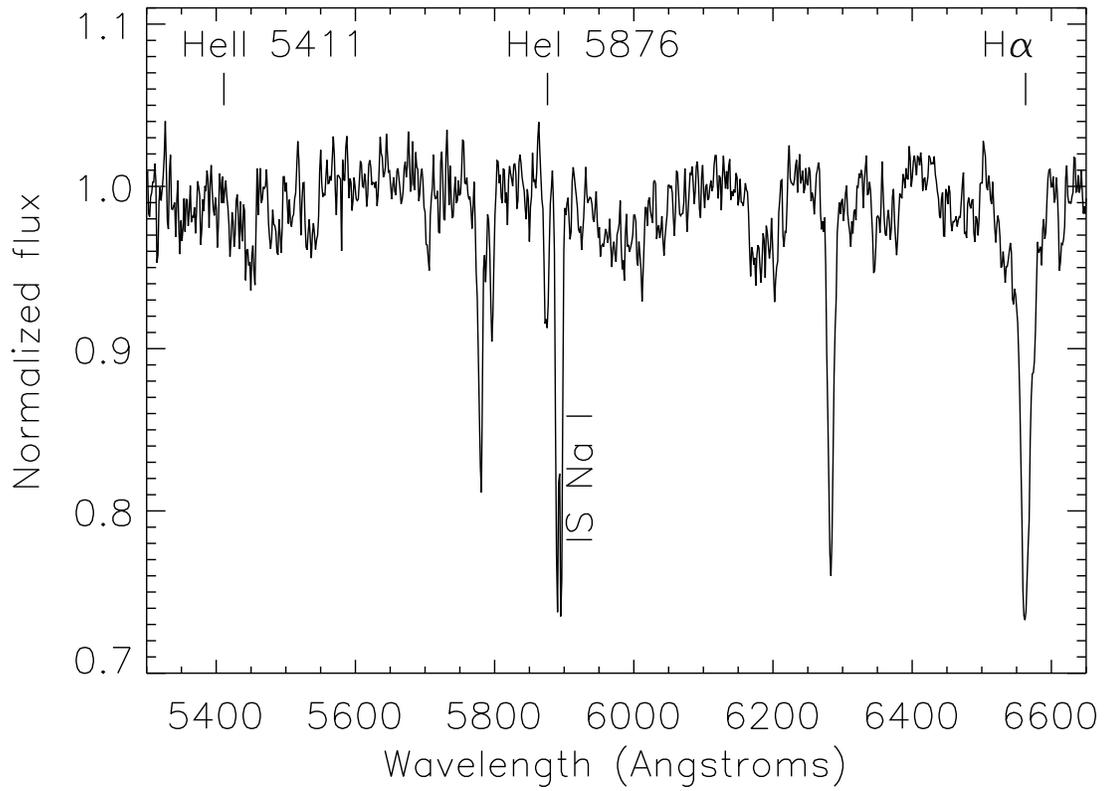}

\caption{Averaged WIRO spectrum of 2MASS03101044+5747035.  Locations
of stellar \ion{He}{2} $\lambda$5411 (not detected), \ion{He}{1}
$\lambda5876$ and  H$\alpha$ are marked.  Other absorption lines are
interstellar, including \ion{Na}{1} $\lambda\lambda$5889,5995 and
several diffuse interstellar band features.     \label{spec} }

\end{figure}

Figure~\ref{spec} shows the average spectrum of Star A between 
5300~\AA\ and 6600 \AA.  The peak signal-to-noise ratio near the
middle of the spectrum is 65:1, but it is significantly lower toward
the blue end beyond the pictured range. Labels denote key spectral
diagnostic lines including \ion{He}{2} $\lambda$5411 (not detected),
and the detected lines of \ion{He}{1} $\lambda$5876 and H$\alpha$.
H$\beta$ and H$\gamma$ are also detected but not labeled owing to
poorer S/N. All other spectral features are interstellar, including
several diffuse interstellar band features \citep[DIB;][]{jenniskens}
and the \ion{Na}{1} doublet, $\lambda\lambda$5889,5895.   The
presence of \ion{He}{1} $\lambda$5876 indicates a relatively hot
star, while the apparent absence of \ion{He}{2} $\lambda$5411 means
that Star A is not among the hottest O stars and is B0.5 or later
\citep{kerton99}.

We used the robust curve fitting package  MPFIT \citep{mpfit} in IDL
to measure equivalent widths and uncertainties for the \ion{He}{2}
$\lambda$5411, \ion{He}{1} $\lambda$5876, and H$\alpha$ lines in this
portion of the spectrum.  Initially line positions, widths, and
depths were allowed to vary as free parameters. \ion{He}{2}
$\lambda$5411 is not detected, so we measured an upper limit by
fixing the width of this line to the value computed for the
$\lambda$5876 line and constraining the position to lie within 1
Angstrom (55 \kms) of the rest wavelength.   Table~\ref{lines} gives
the measured He and H line EWs and uncertainties for Star A.  No
emission lines were detected either at the location of the star or at
locations along the slit covering the bowshock nebula.

\begin{figure}
\plotone{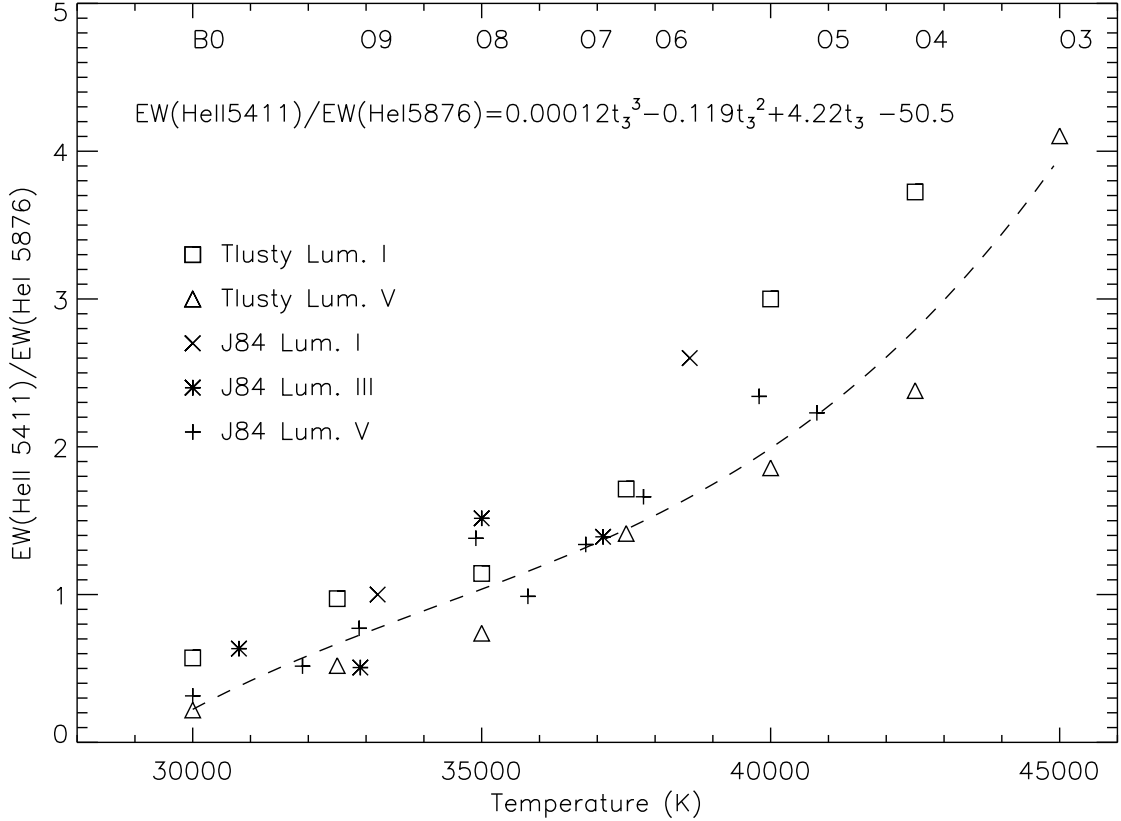}

\caption{Ratio of \ion{He}{2} $\lambda$5411 equivalent width to
\ion{He}{1} $\lambda$58676 EW versus spectral class/temperature for
early type stars from the \citet[][J84]{jacoby84} spectral atlas
(upper ordinate)  and the Tlusty \citep{lanz} model atmospheres
(lower ordinate), for both  luminosity class V (dwarf) and III
(giant) stars, as indicated by the legend. The dashed line shows the
third-order polynomial fit to the luminosity class V models and
spectral atlas stars, as described by the equation.  Star A has a    
ratio EW(HeII $\lambda$5411)/EW(HeI $\lambda$5876)$<0.14$, indicating
a spectral type later than B0.         \label{diag1} }

\end{figure}

The red spectral range covered in our data is not one commonly
employed for spectral typing of hot stars owing to the paucity of
spectral features.  Consequently,  we developed a series of
diagnostic diagrams using the EW of He and H lines to measure the
spectral type of Star A more precisely.  Figure~\ref{diag1} shows the
ratio of \ion{He}{2} $\lambda$5411 equivalent width to that of
\ion{He}{1} $\lambda$5876 as a function of temperature/spectral
type.  Crosses, asterisks, and pluses denote early type stars from
our measurements of spectra in the \citet{jacoby84} spectral atlas
for luminosity classes I, III, V, respectively.  The upper ordinate
gives the spectral type of these stars.  Open squares and triangles
mark the measurements of Tlusty \citep{lanz,hubeny} model atmospheres
with solar metallicity and temperatures as given on the lower
ordinate.  We adopt the relation between spectral type/luminosity
class and effective temperature/surface gravity  provided by
\citet{martins05}  in order to ensure consistency between the upper
and lower ordinates.   The atlas O star dwarfs (pluses) and model O
star atmosphere dwarfs (triangles) show good agreement.  A dashed
line marks a third-order polynomial fit to the pluses and asterisks,
and this expression is given near the top of the plot, where $t_3$ is
stellar effective temperature in thousands K.   Evolved stars and
models with lower surface gravity characteristic of evolved stars lie
systematically above this line.  

\begin{figure}
\plotone{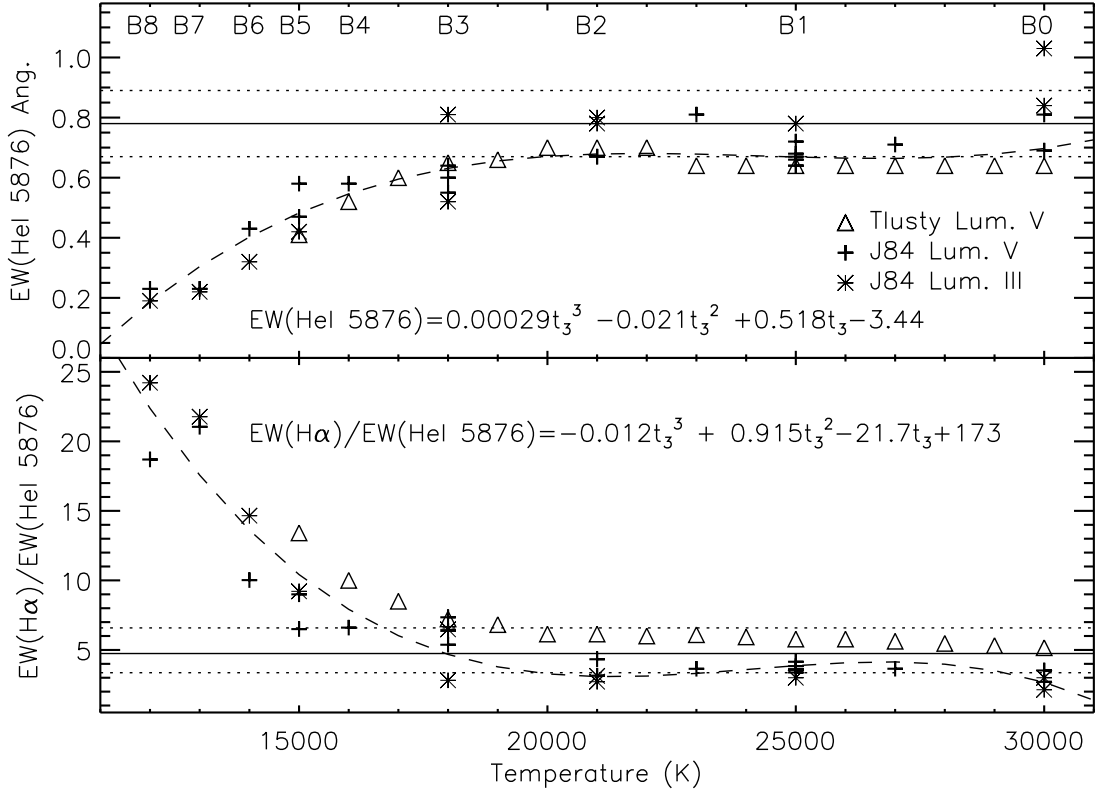}

\caption{({\it upper panel}) EW of \ion{He}{1} $\lambda$5876 versus
temperature/spectral type for B type stars. ({\it lower panel}) Ratio
of H$\alpha$ to \ion{He}{1} $\lambda$5876 equivalent width versus
temperature/spectral type.  Pluses denotes dwarfs (Lum. V) from
Table~\ref{lines} while asterisks denote giants (Lum. III). 
Triangles  are model atmospheres from \citet{lanz07}. The dashed
curve shows a 3rd-order polynomial fit to the pluses and asterisks.  
The solid line and dotted lines show the measurements and 1$\sigma$
uncertainties for Star A.          \label{diag2} } 

\end{figure}

Star A has a ratio EW(\ion{He}{2} $\lambda$5411)/EW(\ion{He}{1}
$\lambda$5876) $< 0.14$, making it cooler than any star on
Figure~\ref{diag1}, and therefore, later than B0.  We constructed a
two additional diagnostic diagrams appropriate for cooler, B-type
stars, using the line measurements from SAO stars in
Table~\ref{lines}, supplemented with a few additional early B star
spectra from the Cygnus OB2 radial velocity survey of
\cite{kiminki07, kiminki09}. He and H line EWs were measured in the
same way using the MPFIT code within IDL.   Figure~\ref{diag2} plots
the EW(HeI $\lambda$5876) versus temperature/spectral type ({\it
upper panel}) and the ratio EW(H$\alpha$)/EW(HeI $\lambda$5876)
versus temperature/spectral type ({\it lower panel}).  As in
Figure~\ref{diag1}, pluses denote dwarfs (Lum. V), asterisks denote
giants (Lum. III), and triangles denote Tlusty \citep{lanz07} model
atmospheres having log(g)=4.0 appropriate to B dwarfs.  Although the
dispersion at any given spectral type is considerable, the data
exhibit a well-defined monotonic trend.  The solid line in each panel
denotes the measurement for Star A, and dotted lines indicate the
1$\sigma$ uncertainties.  The dashed curve is a 3rd-order polynomial
fit to all of the pluses and triangles. Analytic expressions for
these curves appear within each panel, where $t_3$ is the temperature
in thousands K. We match temperature to spectral type using the
relations of \citet{schmidt-kaler82} which are very similar to the
calibration of \citet{underhill79}.  While the measurements on real
stars and model atmospheres compare favorably in the upper panel, the
models are systematically displaced toward larger 
EW(H$\alpha$)/EW(HeI $\lambda$5876) ratios in the lower panel.   We
believe that this is the result of broad Balmer wings which are
measurable in the model H$\alpha$  lines but not in the lower S/N
data.  Interestingly, the  giants and dwarfs show good agreement in
both panels.  

Figure~\ref{diag2} shows that the spectral type of Star A is near
B0.5 on the basis the EW(HeI $\lambda$5876) measurement ({\it upper
panel}) and nearer B3 on the basis of the EW(H$\alpha$)/EW(HeI
$\lambda$5876) measurement ({\it lower panel}).   Unfortunately,
there are few spectral diagnostics sensitive to temperature for early
B stars in the spectral range available.  Traditional spectral type
diagnostics in the blue end of the optical regime are not helpful in
our spectra because of low S/N---a consequence of reduced
instrumental sensitivity and high interstellar reddening which
suppresses the blue flux from Star A.    We adopt a spectral type of
B1.5$\pm$1.0.   H or He emission lines, often seen in B supergiants,
are not evident in our spectrum of Star A.  Furthermore, the
H$\alpha$ and H$\beta$ linewidths in star A are 15--18\AA FWHM---
more typical of dwarfs and much broader than the Balmer lines in any
of our comparison B giants.  Therefore, we adopt a luminosity of
class of V.

\subsection{Star A Spectral Energy Distribution and Distance}

Table~\ref{sed.tab} lists eight optical and infrared photometric
measurements for Star A compiled from available databases.  These
include the B2, R2, I2 entries from the USNO-B1.0 catalog
\citep{monet}, the JHK entries from the 2MASS Catalog, and \SST\
[3.6] and [4.5] data from the GLIMPSE Point Source Catalog. 
Magnitudes have been converted to Jy using standard zero points
available in the respective documentation.  The optical data points
are particularly uncertain,  as they come from photographic surveys,
so we adopt uncertainties of 15\%. The infrared photometry from 2MASS
and GLIMPSE has uncertainties of 0.03 mag, or about 3\%.  The red
colors in any two bandpasses suggest a high degree of reddening for
an early B star which has close to zero intrinsic color.  

\begin{figure}
\plotone{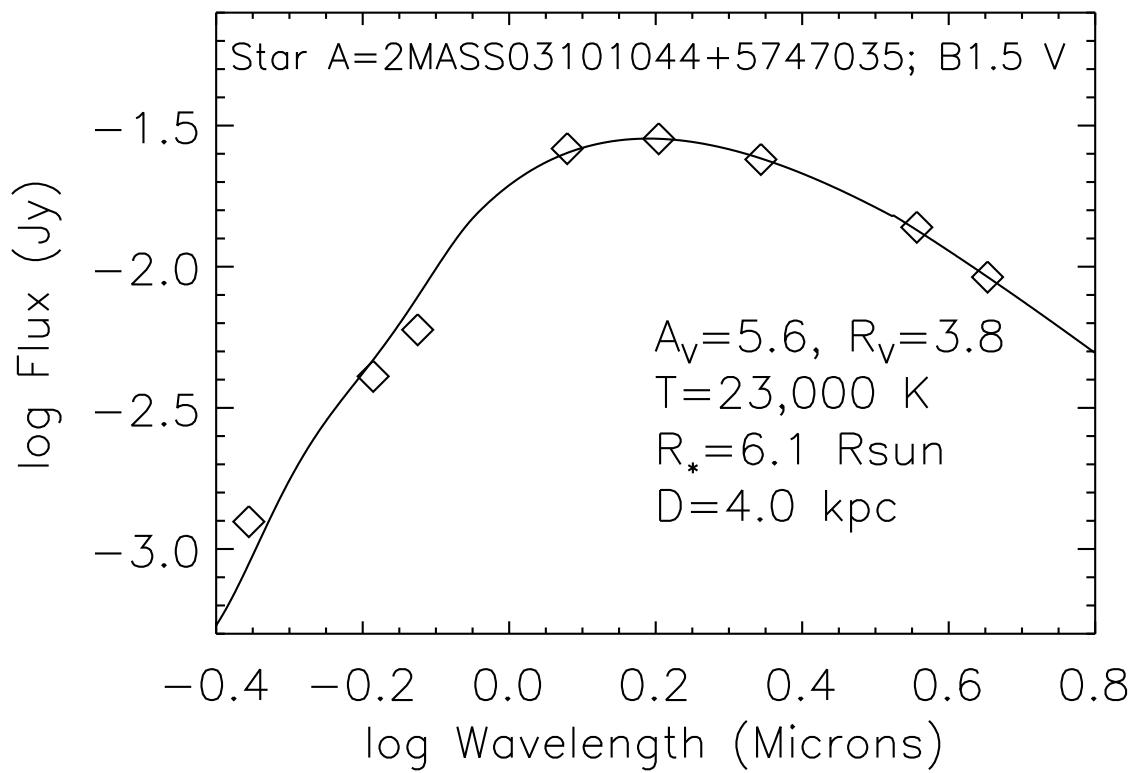}

\caption{Spectral energy distribution for Star A.  The curve is a
blackbody with T=23,000K and radius R=6.1 R$_\odot$ appropriate for a
B1.5~V star, interpolated from the data of \citet{drilling}, with
$A_V=5.4$ mag and a \citet{ccm} extinction law in the optical and
\citet{indebetouw07} extinction law in the mid-IR. A distance of 4.0
kpc yields a good fit. \label{sed} }

\end{figure}

Figure~\ref{sed} shows the optical and infrared spectral energy
distribution of Star A.  The solid curve is a blackbody approximation
to a B1.5~V star with temperature T=23,000 K, radius R=6.1~R$_\odot$
\citep[interpolated from][]{drilling}, and $A_V$=5.6 mag of reddening
according to a \citet{ccm} extinction curve in the optical and
near-infrared and an \citet{indebetouw07} reddening law in the
mid-infrared. The data are well fit by the model for a distance of
$\sim$4.0 kpc.  The fit is quite good in the infrared where the
photometry is most accurate and less good in the optical regime where
photomteric uncertainties are  considerable.  Nevertheless, the SED
constrains the extinction with within about ${\Delta}A_V\approx0.4$. 
No infrared excess is apparent, as might be expected for evolved
massive stars having free-free emission or stars hosting
circumstellar material. This is consistent with our adoption of a
dwarf (i.e., main sequence) luminosity class.  Based on the fit in
Figure~\ref{sed}, we adopt a distance of 4.0$\pm$1 kpc to Star A and
the bowshock nebula, where the uncertainty is dominated by the
uncertainty of the star's spectral type and intrinsic luminosity.  At
this distance 1\arcsec\ corresponds to  0.019 pc and the
$\sim$2\arcmin\ angular diameter of the bowshock nebula corresponds
to 2.3 pc. This also means that the 7\arcsec\ angular distance  from
Star A to the bowshock apsis translates to a linear distance of
$\geq$0.13 pc, where the inequality accounts for the unknown
inclination relative to the line of sight.  Given that bowshock
nebulae are most easily recognized  when seen edge-on, the
inclination is likely to be near 90\degr\ and the standoff distance,
$R_0 \simeq 0.13$ pc, similar to  bowshocks from stars \#2 and \#5 in
the \citet{kobulnicky10} study  of Cygnus~OB2. 

\subsection{Identification of $^{12}$CO Molecular Complex}
  
\begin{figure}
\plotone{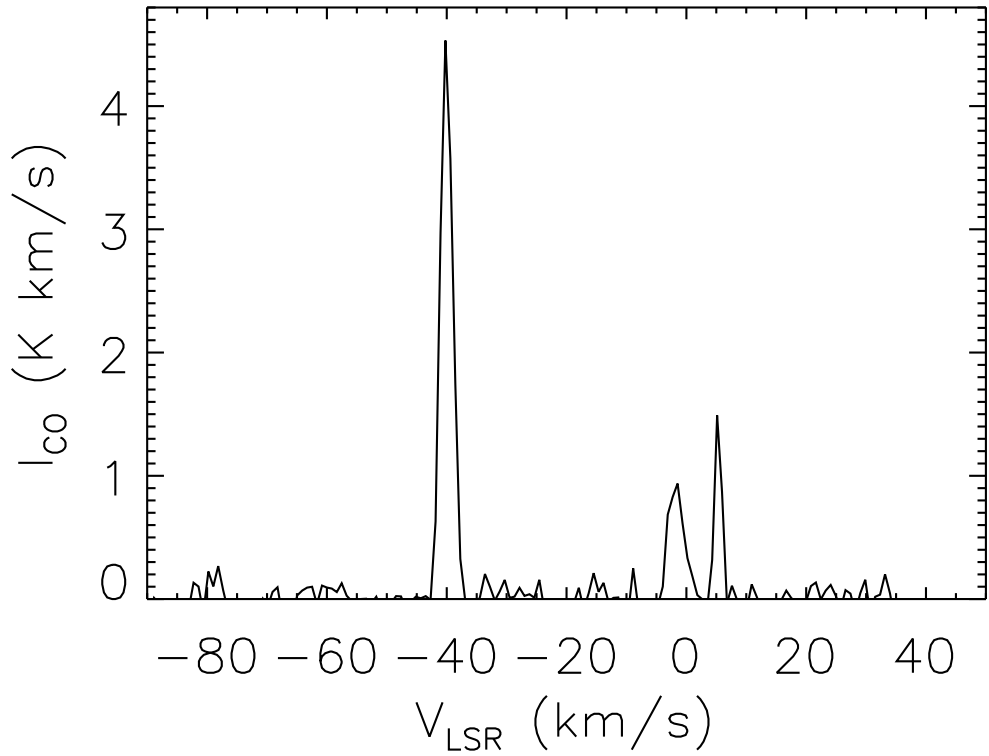}

\caption{A 1-dimensional \tco\ spectrum from the CGPS along the
IRAS~03063+5735  sightline averaged over a square 2\arcmin\ region.  
The molecular component near -40 \kms\ bears a strong morphological 
similarity to the diffuse emission from warm dust.   \label{1Dco} }

\end{figure}

\begin{figure}
\plotone{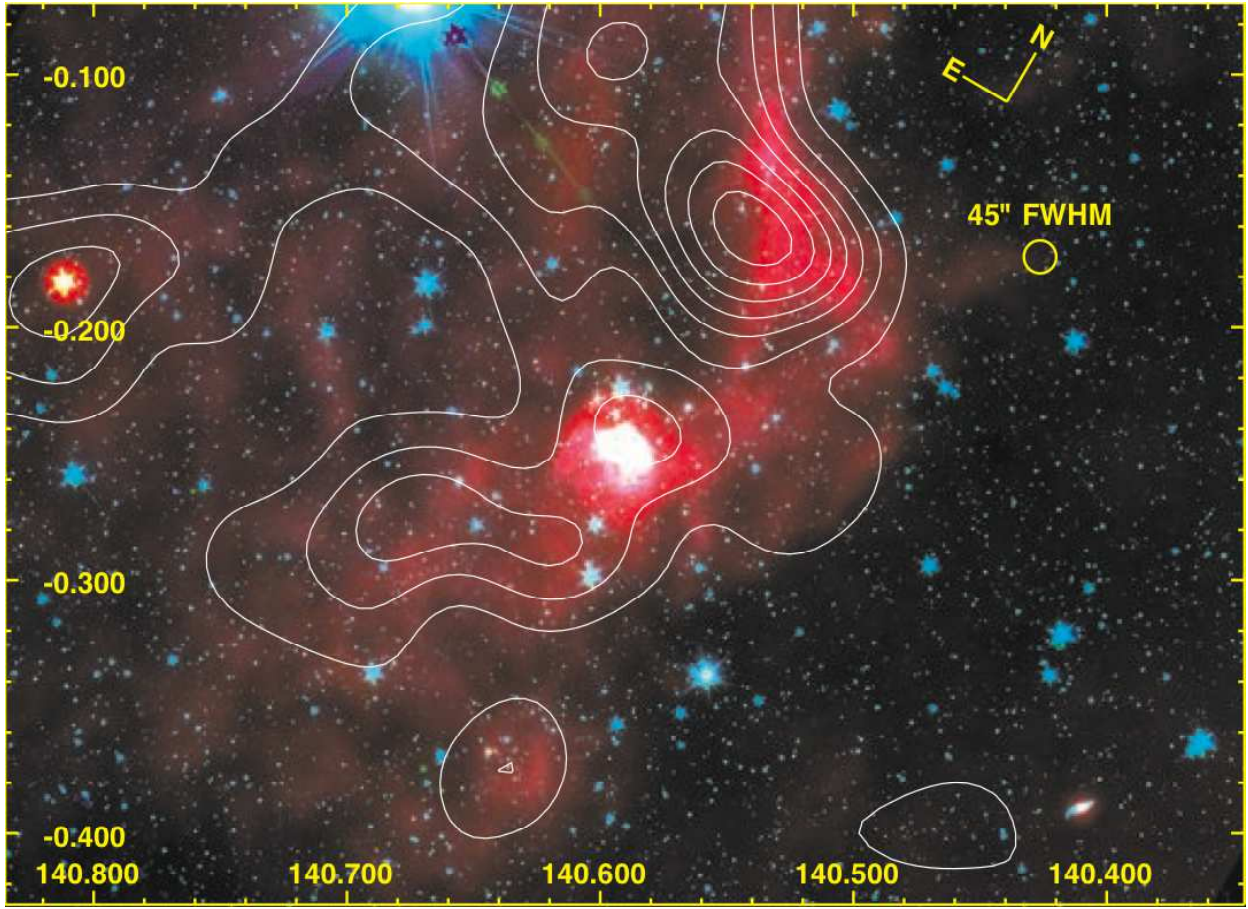}

\caption{A color representation of the larger IRAS~03063+5735 region
in  Galactic coordinates with \SST\ GLIMPSE [3.6] in blue, [4.5] in
green, and  the \emph{WISE} [22] in red.  Contours show the $^{12}$CO
brightness map  from the CGPS integrated over the LSR velocity range
-38 to -42 \kms.  The  $^{12}$CO contours correspond to $H_2$ column
densities of 1.5 $\times10^{21}$ cm$^{-2}$ -- 10.5 $\times10^{21}$
cm$^{-2}$ in  steps of 1.5$\times10^{21}$ cm$^{-2}$.   \label{comap}
}

\end{figure}

We examined the \tco\ velocity cube of the IRAS~03063+5735 region
from the Canadian Galactic Plane Survey \citep{cgps}.  The data cube
has velocity channels of width 0.81 \kms\ and a beamsize of
45\arcmin\ FWHM.  There are three peaks of molecular emission along
this sightline.  Figure~\ref{1Dco} shows the 1-dimensional \tco\
spectrum summed over the 2\arcmin~square region surrounding
IRAS~03063+5735. The first peak near 0~\kms\ covers this entire
region out to 10\arcmin\ from the bowshock nebula with uniform
surface brightness.  This molecular component is likely to be
affiliated with very local gas. The second component near $+5$~\kms\
appears as a small filament that crosses near the bowshock but bears
no morphological resemblance to the diffuse ISM emission seen in the
\SST\ and {\it WISE} infrared images.  The third component centered near
$-40$~\kms\ shows a strong channel-by-channel morphological
resemblance to the mid-IR PAH and hot dust emission. 
Figure~\ref{comap} shows a color view of the extended IRAS~03063+5735
region with \SST\ [3.6] and [4.5] in blue and green, respectively,
and the {\it WISE}  [22] in red. Contours show the \tco\ intensity
integrated over the $-38$ -- $-42$~\kms\ velocity channels.  The
\tco\ contours correspond to intensities of 5 through 35 K \kms\ in
steps of 5~K, equivalent to H$_2$ column densities of 1.5 through
10.5 $\times10^{21}$ cm$^{-2}$ adopting a standard conversion factor
of $\alpha$=3.0$\times10^{20}$ cm$^{-2}$ (K \kms)$^{-1}$
\citep{solomon87}.  Note that the location of the bowshock nebula
coincides with a local maximum in the H$_2$ column density of
approximately 4.5 $\times10^{21}$ cm$^{-2}$.  If we estimate that
this feature is a roughly spherical molecular cloud with radius
$r=1$\arcmin\ $\equiv$  1.10 pc, then the implied number density is
$N_{H2}\simeq 600$ cm$^{-3}$.  We caution that this is only a gross
average, as \tco\ is known to freeze out onto dust grains in the cold
dense cloud interiors.  

The similarity of the \tco\ and diffuse ISM morphology, along with
the identification of a local \tco\ peak at the bowshock location,
supports the association of the bowshock with the molecular complex.
The radial velocity of Star A, which we measure to be
$-59\pm18$~\kms\  by a Gaussian fit to the \ion{He}{1} $\lambda$5876
line, adopting a rest wavelength of  5875.6 \AA.  This lends further
support to the association of the bowshock with this particular
molecular complex.  In principle, one might expect a runaway star to
show anomalously large velocities relative to the surrounding medium.
However, given the clear bowshock morphology of the associated
nebula, we would also expect such objects to have velocity vectors
predominantly in the plane of the sky.  Hence, the similarity between
the radial velocity of Star A and the  molecular cloud upon which it
impinges is consistent with the observed geometry.   

The \citet{brandblitz93} rotation curve in the outer Galaxy shows
that radial velocity is not a very precise distance indicator along
this $\ell=$140\degr\ sightline.  Their Figure~2b shows gas at
$\sim -40$ \kms\ associated with a wide range of distances from 2 to
6 kpc. Our spectrophotometric distance estimate to Star~A of
4.0$\pm$1 kpc  links the molecular gas at $-40$~\kms\ to a distance
comfortably within that range.   

\subsection{IRAS~03063+5735 at other Wavelengths}

IRAS~03063+5735 is not detected to a 3$\sigma$ limit of 0.037 Jy in
the CGPS 1420 MHz radio map of this region. \citet{wouterloot93}
searched for but did not detect any OH or water masters from this
infrared source.     

As first demonstrated by \citet{rub68} the observed flux density of
radio emission at frequency $\nu$ ($F_\nu$) from an optically-thin,
isothermal, \ion{H}{2} region at a known distance can be related to
the H-ionizing photon flux ($Q$) from the exciting OB star(s). In
this study we will use values of $Q$ tabulated from model atmospheres
of OB stars to determine if an \ion{H}{2} region surrounding an early
B-type star should be visible in our radio continuum images.

\citet{kerton99} present a version of the\citet{rub68} formula for
observations at 1420 MHz of an isothermal  (7500 K) \ion{H}{2} region
containing only hydrogen. Using the stellar calibration of
\citet{crowther05} we find a B1.5~V star has $\log(Q) = 46.0$ which
corresponds to F$_{1420} = 900 d^2 f$, where $d$ is the distance in
kpc and $f$ is essentially a covering factor (ie., $f=1$ would
correspond to an ionization bounded region, $f=0.25$ could apply to a
blister-type region).  For $f=1$ and $d=4$ we find F$_{1420}=7$ mJy.
This corresponds to a source with $T_B = 1.1$~K in the CGPS ignoring
any beam dilution. The average and standard deviation of brightness
temperature of our 1420 MHz image in a 10x10 arcmin box surrounding
the position of the bowshock is $T_B = 5.68\pm0.08$~K so the distance
and spectral type are consistent with the non-detection of this
object in the CGPS radio contiuum images. The non-detection in the
CGPS image actually puts a fairly strong limit on how early the star
could be. For example, at O9~V $\log(Q)=48.1$, and repeating the
above calculation leads to $T_B = 139$~K. Even with a very low value
for $f$ the \ion{H}{2} region associated with a late O type star
would be easily detectable. All of this is consistant with our
non-detection of the \ion{He}{2} $\lambda$4511 line in the optical
spectrum which shows that Star A is B0.5~V or later.

\subsection{Bowshock SED}

\begin{figure}
\plotone{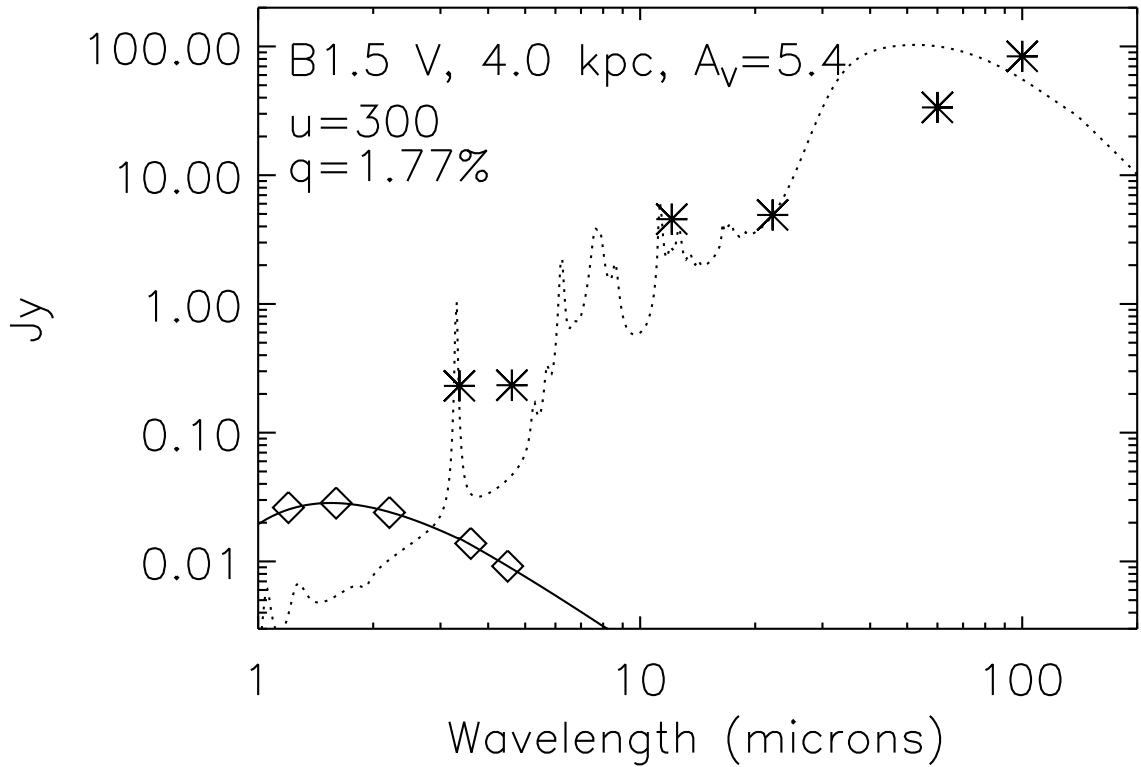}

\caption{The spectral energy distribution of the IRAS~03063+5735 
bowshock nebula (asterisks) from the four \emph{WISE} bandpasses and 
two \emph{IRAS} measurements at 60 and 100 $\mu$m.   Diamonds and the
solid line show the extincted  stellar SED.  The dotted line shows a
\citet{draineli07} dust model with radiation intensity $u=300$ times
the mean interstellar radiation field \citep{mmp83} and PAH
contribution 1.77\%. \label{BSsed} }

\end{figure}

Figure~\ref{BSsed} shows the (background-subtracted) spectral energy
distribution of the nebula (asterisks) as measured in a large
lima-bean shaped aperture in the {\it WISE}  bandpasses at 3.6, 4.5, 
12 and 22 $\mu$m.  The star is detected in the {\it WISE}  W1 and W2
bands, so we have used the higher-resolution \SST\ GLIMPSE photometry
to subtract off the star's contribution to the {\it WISE}  W1 and W2
measurements.   The diamonds and solid line show the data and
extincted blackbody fit for Star A alone.  The dashed line shows a
\citet{draineli07} dust model fit by eye for  a radiation field
$u=300$ times the average interstellar radiation field of
2.17$\times10^{-2}$ erg~s$^{-1}$ cm$^{-2}$ \citep{mmp83}, a PAH
fractional contribution of 1.77\% and a mean  ISM number density of
600~ cm$^{-3}$.  The {\it WISE}  W2 [4.5] datum lies well above the
model, possibly because of a contribution from shocked molecular
hydrogen in this band.  The \emph{IRAS} 60 and 100 $\mu$m data also
show relatively poor agreement with the model, but the uncertainties
on these data are large owing to the large (4\arcmin) beamsize and
background subtraction complications.    The diffuse infrared surface
brightness varies substantially over the  nebula, suggesting
substantial spatial variations in the local dust and gas density. 
Variations in temperature and density throughout the nebula are
certain to render the single averaged SED a gross
oversimplification.   Hence, the dotted line fit is mostly for
illustrative purposes, especially given large uncertainties on the
local number density in the vicinity of the bowshock.

\subsection{Search  for Associated Stellar Cluster}

The apparent increase in stellar density just to the north-east of
the Star A initially suggested the possibility of a stellar cluster
at this location. However, a color-magnitude diagram using JHK
photometry of 2MASS  Catalog sources within a 1\arcmin\ radius shows
that the stars have a considerable spread in color (hence,
reddening), and no clustering in either color-color or
color-magnitude diagrams is evident. Therefore, we find no evidence
of a stellar cluster in this vicinity, at least to the depths  probed
by the 2MASS and GLIMPSE catalog data.  

\section{Discussion}

\subsection{Runaway or Low-Velocity Star?}

The expected proper motion of an object at 4.0 kpc distance with a
transverse motion of 30 \kms\ is $\sim$1 mas yr$^{-1}$, too small to
have a measured proper motion even over decade time scales in
multi-epoch photographic plate surveys.  Indeed, the USNO-B1.0
Catalog does not report a proper motion.  Hence, there is no direct
evidence for Star A having a high space velocity.  Rather, it is
inferred as a possibility from the morphology of the nebula.  

It is also possible that low-velocity stars moving in a medium with a
density gradient \citep{wilkin} can produce a bowshock-like nebula.
The proximity of Star A to a molecular cloud having a maximum column
density just ahead of the bowshock (see Figure~\ref{comap}) makes this
explanation worth considering.  Perhaps Star A formed within or near
the pictured molecular complex.  However, formation of such a massive
star typically requires relatively high column densities
characteristic of massive molecular clouds and would be accompanied by
the formation of other lower mass stars comprising a small embedded
cluster.  We find no evidence of either high column density molecular
clouds (the highest is 10.5$\times10^{21}$ \cmmb\ at 85\arcsec\
$\equiv$1.6 pc north of the bowshock) or a stellar cluster containing
lower mass stars.  Furthermore, the observed excess emission in the
[4.5] band (Figure~\ref{BSsed}) would be well explained by emission
from shocked $H_2$ if the nebula were a genuine shock. It is not clear
whether a low-velocity B1.5 star carving a cavity within a local
density gradient would generate a wind shock sufficient to produce
excitation of $H_2$.  More detailed spectral information and shock
modeling would be needed to determine if this is a possibility.  Given
the current data, we prefer the interpretation that Star~A is
encountering the molecular cloud for the first time after having
traveled some distance from its point of formation at a high velocity.

\subsection{Possible Origins of a Runaway Star}

The orientation of the cometary bowshock nebula suggests that Star
A's direction of travel is roughly toward the equatorial north.  We
searched the SIMBAD database for young stars, star clusters, and
\HII\ regions up to 3 degrees to the south to identify candidate
birthplaces of Star A.  At a distance of 4.0 kpc and an assumed
tangential velocity of 30 \kms\ (30 pc Myr$^{-1}$) in the plane of
the sky, it would be reasonable to suppose that the star has
traversed up to 600 pc (8\degr on the sky ) or even more during its
few$\times$10 Myr main-sequence lifetime. However,  given that the
star is already below the Plane and moving toward Galactic north,
adopting the maximum travel time allowed by the main-sequence
lifetime  implies a formation locale well outside the scale height of
Milky Way molecular clouds required to form such a massive star. 
Therefore, we prefer the interpretation that the runaway has been
ejected recently and confine our search within several degrees of its
current location.   

\begin{figure}
\plotone{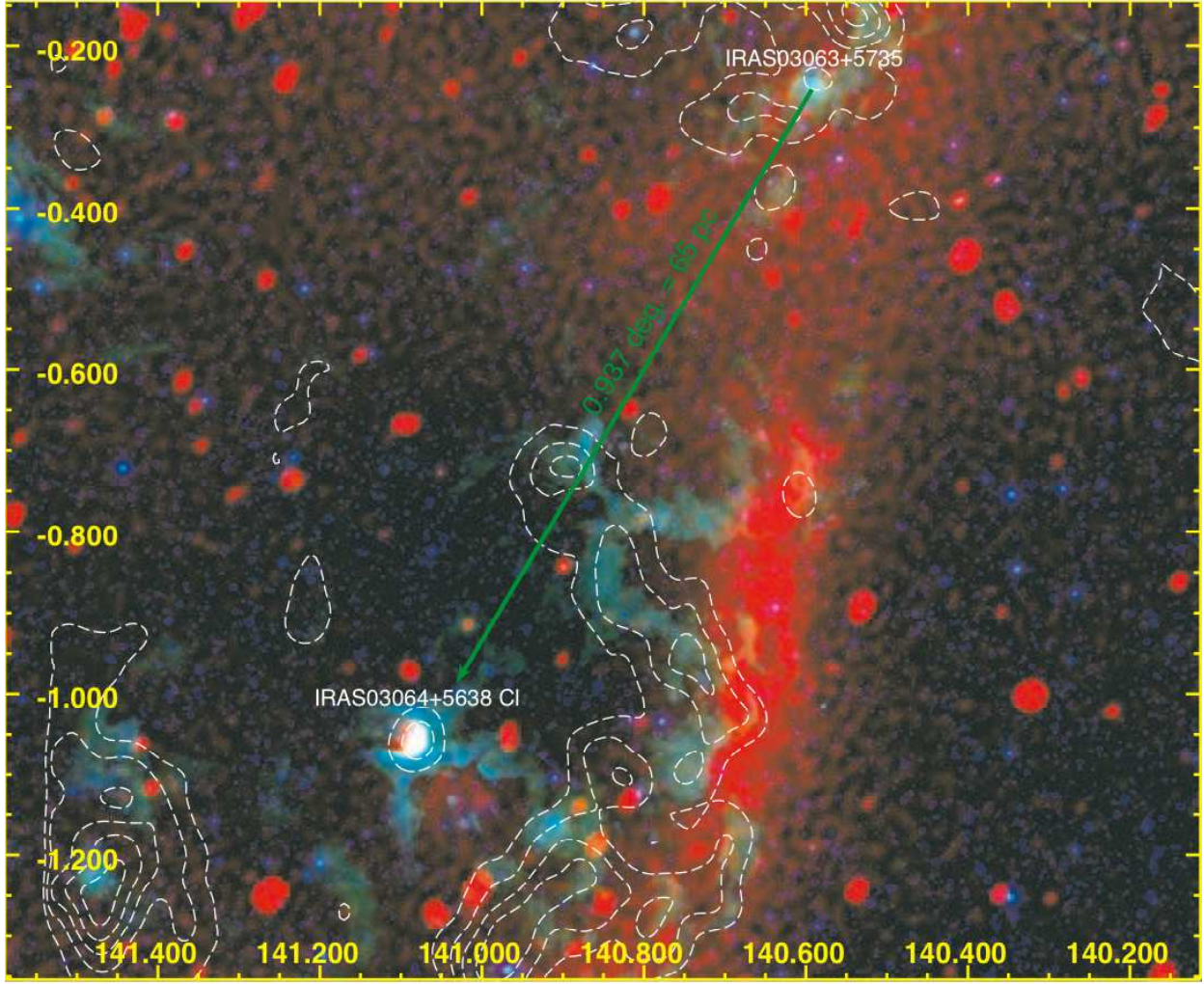}

\caption{Color representation of the field around IRAS~03063+5735 as
seen in the 1420 MHz CGPS radio 
continuum map (red), \emph{WISE} W3 [11.1] (green), and \emph{WISE} 
W2 [4.5] (blue).  Contours depict the \tco\ integrated between LSR
velocities of -31 \kms\ and -43 \kms.   A green arrow points rearward
along the bowshock axis of symmetry to a hypothetical birthplace of
Star A near the \HII\ region and stellar cluster associated with
IRAS~03064+5638 at a projected distance of 65 pc.    \label{wide} }

\end{figure}

One likely candidate for Star A's birthplace is the \HII\ region and
stellar cluster associated with IRAS~03064+5638 located 0.94\degr\
(64 pc) due south.  Figure~\ref{wide} shows a three-color image of
the field around Star A with the CGPS 1420~MHz radio continuum in
red, the {\it WISE}  [11.1] in green, and the {\it WISE}  [4.5] in blue. 
Contours indicated the \tco\ brightness of 5, 10, 15, 25, 30 K~\kms\
integrated between LSR velocities of $-31$ and $-43$~\kms.  The
bright ridge of radio emission extending 2\degr\ north to south is a
thermal radio filament at an unknown distance and powered by a
yet-unidentified source \citep{green89}.  Both the bowshock nebula
IRAS~03063+5735 and IRAS~03064+5638 have \tco\ peaks in this velocity
range, suggesting a common distance.  IRAS~03064+5638 contains a
stellar cluster  characterized by \citet{carpenter93} as having a few
mid-B type stellar members, for their adopted distance of 2.2 kpc. 
If located at 4.0 kpc instead, these become early B stars of
comparable mass and age to Star A.  The 1420 MHz radio flux of
IRAS~03064+5638 is $101\pm3$~mJy, equivalent to a B0~V star at the
adopted distance.   Apparently, the IRAS~03064+5638 cluster contains
at least one star more massive than Star A, making it an environment 
from which Star A could plausibly have been ejected.  Three-body
encounters which eject one star often result in the formation of  a
tight binary.  Therefore, our working scenario predicts the existence
of close early-B binary within IRAS~03064+5638.  However, the
extinction appears high at this location, and no suitably bright
optical sources are visible.  

The peak molecular column density at  IRAS~03064+5638  is
approximately $N(H_2)$=4.5$\times10^{21}$ \cmmb.   The adjoining molecular
clouds which lack bright IR emission have peak $N(H_2)$ almost twice
this value, suggesting other possible sites of intermediate-mass star
formation nearby. These, yet unidentified, sites of star
formation could have also been the birthplace of star A.

Another possible origin for Star A is the open cluster  NGC~1220,
located a full 3\degr\ ($\sim$201 pc) to the south of Star A, but
along the same vector pictured in Figure~\ref{wide}.
\citet{ortolani02} identified 26 probable members of this open
cluster, estimating a distance of 1800$\pm$200 pc and an age of
60~Myr.  While this denser and more populous cluster provides a
suitable environment for ejecting a B star via N-body encounters, the
much smaller heliocentric distance  (if correct) makes a physical
connection between NGC~1220 and IRAS~03063+5735/Star A less likely.

\section{Conclusions}

We have identified an \emph{IRAS} infrared source IRAS~03063+5735,
previously reported as a galaxy, to be a bowshock nebula powered an
early B-type (B1.5 V) star. The infrared source was selected from
among several hundred sources having \emph{IRAS} color-color criteria
matching those expected from soft UV radiation fields generated by
intermediate-mass (i.e., B-type) stars.  Our measurement of the $-58$
\kms\ LSR radial velocity of the star, combined with \tco\ line maps,
allow us to identify a molecular cloud complex at a common velocity
and $\sim$4.0 kpc distance.  The bowshock is plausibly produced by the
supersonic motion of a runaway B star as it impinges upon a local
maximum in the molecular column density of $\sim4.5\times10^{21}$
\cmmb.  However, a low-elocity star within a local density gradient
may also be capable of producing the observed phenomena.  We identify
the \HII\ region and stellar cluster associated with IRAS~03064+5638
at a projected distance of 64 pc as one plausible birth site for a
runaway star.  The association of a star having a spectrophotometric
distance with a particular molecular feature adds another data point
to help anchor the (poorly defined) rotation curve in the outer
Galaxy.

\acknowledgments

We thank WIRO staff James Weger and Jerry Bucher for their
indefatigable efforts that enable the Observatory to produce science.
We thank Ed Churchwell, Bob Benjamin, and an anonymous referee for
helpful suggestions.  Grants from the National Science Foundation
through AST-09-08239 and NASA through 09-ADP09-0097 enabled this
work. This publication makes use of data products from the Wide-field
Infrared Survey Explorer, which is a joint project of the University
of California, Los Angeles, and the Jet Propulsion
Laboratory/California Institute of Technology, funded by the National
Aeronautics and Space Administration. This publication makes use of
data products from the Two Micron All Sky Survey, which is a joint
project of the University of Massachusetts and the Infrared Processing
and Analysis Center/California Institute of Technology, funded by the
National Aeronautics and Space Administration and the National Science
Foundation. This research has made use of the SIMBAD database,
operated at CDS, Strasbourg, France

{\it Facilities:} \facility{WIRO ()}, \facility{Spitzer () }, \facility{WISE () }

{}

\clearpage

\begin{deluxetable}{lccccc}
\tablecolumns{6}
\tablewidth{0pc}
\tablecaption{Star A and Comparison Star Line Measurements \label{lines}}
\tablehead{
\colhead{Star} &
\colhead{EW(HeII $\lambda$5411)} &
\colhead{EW(He\sc{I} $\lambda$5876)} &
\colhead{EW(H$\alpha$)} &
\colhead{Spectral Type} & 
\colhead{Ref.}       \\
\colhead{} &
\colhead{(\AA)} &
\colhead{(\AA)} &
\colhead{(\AA)} &
\colhead{} &
\colhead{} }
\startdata
Star A         & 0.11 (0.10) & 0.78 (0.11) &  3.70 (0.71) & B1.5~V     & 1 \\
\hline
DL Cam         & 0.24 (0.01) & 1.03 (0.06) &  2.20 (0.34) & B0~III   & 2 \\
HD36822        & 0.13 (0.01) & 0.84 (0.01) &  2.52 (0.31) & B0~III   & 2 \\
HD23180        & $<$0.01     & 0.78 (0.03) &  2.34 (0.43) & B1~III   & 2 \\
HD30836        & $<$0.01     & 0.80 (0.02) &  2.17 (0.24) & B2~III   & 2 \\
HD29248        & $<$0.01     & 0.78 (0.02) &  2.44 (0.31) & B2~III   & 2 \\
HD31237        & $<$0.01     & 0.81 (0.02) &  2.28 (0.27) & B3~III   & 2 \\
$\epsilon$ Cas & $<$0.01     & 0.52 (0.01) &  3.37 (0.52) & B3~III   & 2 \\
$\delta$ Per   & $<$0.01     & 0.42 (0.01) &  3.87 (0.51) & B5~III   & 2 \\
$\lambda$ Cet  & $<$0.01     & 0.32 (0.01) &  4.69 (0.68) & B6~III   & 2 \\
HD16727        & $<$0.01     & 0.22 (0.01) &  4.79 (0.74) & B7~III   & 2 \\
$\gamma$ Per   & $<$0.01     & 0.19 (0.01) &  4.60 (0.65) & B8~III   & 2 \\  
\hline
MT378          & $<$0.09     & 0.81 (0.05) &  2.19 (0.14) & B0~V     & 3 \\
MT429          & 0.13 (0.03) & 0.69 (0.02) &  2.43 (0.18) & B0~V     & 3 \\
HD22951        & 0.02 (0.01)  & 0.71 (0.02) &  2.59 (0.32) & B0.5~V   & 2 \\
MT605          & 0.05 (0.02) & 0.64 (0.02) &  2.31 (0.15) & B1~V     & 3 \\
MT365          & 0.05 (0.02) & 0.72 (0.03) &  2.43 (0.27) & B1~V     & 3 \\
MT605          & 0.05 (0.02) & 0.64 (0.02) &  2.31 (0.15) & B1~V     & 3 \\
MT187          & 0.03 (0.02) & 0.67 (0.02) &  2.39 (0.16) & B1~V     & 3 \\
HD21856        & $<$0.01     & 0.68 (0.02) &  2.61 (0.33) & B1~V     & 2 \\
HD24131        &$<$ 0.01     & 0.66 (0.01) &  2.74 (0.36) & B1~V     & 2 \\
HD11241        &$<$0.01      & 0.81 (0.02) &  2.95 (0.56) & B1.5~V   & 2 \\
MT234          &$<$0.01      & 0.67 (0.03) &  2.90 (0.27) & B2~V     & 3 \\
HD23990        &$<$0.01      & 0.67 (0.01) &  2.11 (0.41) & B2~V     & 2 \\
MT298          &$<$0.01      & 0.64 (0.07) &  3.44 (0.54) & B3~V     & 3 \\
29 Per         &$<$0.01      & 0.60 (0.01) &  3.82 (0.56) & B3~V     & 2 \\
34 Per         &$<$0.01      & 0.55 (0.01) &  4.04 (0.63) & B3~V     & 2 \\
53 Per         &$<$0.01      & 0.58 (0.01) &  3.83 (0.54) & B4~IV    & 2 \\
31 Per         &$<$0.01      & 0.58 (0.02) &  3.77 (0.65) & B5~V     & 2 \\
HD21278        &$<$0.01      & 0.47 (0.01) &  4.23 (0.63) & B5~V     & 2 \\
HD24504        &$<$0.01      & 0.43 (0.02) &  4.31 (0.76) & B6~V     & 2 \\
HD18537        &$<$0.01      & 0.23 (0.01) &  4.84 (0.80) & B7~V     & 2 \\
$\beta$ Per    &$<$0.01      & 0.23 (0.01) &  4.30 (0.65) & B8~V     & 2 \\
\enddata
\tablecomments{1--This work; 2--SAO Star Catalog; 3--\citet{MT91,kiminki07} }
\end{deluxetable}
\clearpage

\begin{deluxetable}{ccrc}
\tablecolumns{4}
\tablewidth{0pc}
\tablecaption{2MASS 03101044+5747035 Photometry \label{sed.tab}}
\tablehead{
\colhead{Wavelength} &
\colhead{Mag.} &
\colhead{Flux} &
\colhead{Source} \\
\colhead{($\mu$m)} &
\colhead{} &
\colhead{(mJy)} }
\startdata
0.44  & 16.33 & 1.25 & 1 \\
0.65  & 14.54 & 4.09 & 1 \\
0.75  & 13.93 & 5.68 & 1 \\
1.2   & 12.04 & 26.2 & 2 \\
1.6   & 11.38 & 28.4 & 2 \\
2.2   & 11.03 & 24.0 & 2 \\
3.6   & 10.82 & 13.8 & 3 \\
4.5   & 10.78 & 9.19 & 3 \\
\enddata
\tablecomments{1--The USNO-B1.0 Catalog, \citep{monet}; 2--2MASS \citep{2MASS}; 
3--GLIMPSE~360 Point Source Catalog}
\end{deluxetable}
\clearpage

\begin{deluxetable}{ccc}
\tablewidth{0pc}
\tablecolumns{3}
\tablecaption{IRAS~03063+5735 Photometry \label{BSsed.tab}}
\tablehead{
\colhead{Wavelength} &
\colhead{Flux} &
\colhead{Notes}  \\
\colhead{($\mu$m)} &
\colhead{(Jy)} &
\colhead{} }
\startdata
W1 3.6  & 0.21  &  1 \\
W2 4.5  & 0.23  &  1 \\
W3 12.1 & 4.6   &  1 \\
W4 22.0 & 4.9   &  1 \\
IRAS 60 & 33.7  &  2 \\
IRAS100 & 83.8  &  2 \\
\enddata
\tablecomments{We adopt a 10\% uncertainty on the flux measurements at all {\it WISE}  bandpasses
and a factor of 2 for the two \emph{IRAS} measurements.
1--The reported flux reflects the entire nebula in a large aperture
minus the flux from the star, as measured from \SST\ GLIMPSE photometry, as reported
in Table~\ref{sed.tab}; 2--\emph{IRAS} Point Source Catalog value.  The \emph{IRAS} measurements at 12 and 25 
$\mu$m  are systematically a factor of about 1.8 lower than the W3,W4 {\it WISE}  measurements in 
similar bandpasses.  This discrepancy probably reflects the much larger \emph{IRAS} beamsize
and associated uncertainty in subtracting the diffuse background emission.  }
\end{deluxetable}

\clearpage

\clearpage

\clearpage

\clearpage

\clearpage

\clearpage

\clearpage

\clearpage

\clearpage

\end{document}